\documentclass{ws-procs9x6}

\usepackage{graphicx}
\usepackage[english]{babel}
\usepackage{color}
\usepackage{amsmath,amssymb,amsfonts}

\newcommand{\ener}[1]{ \varepsilon_{#1} }
\newenvironment{eq}{\begin{eqnarray}}{\end{eqnarray}}
\newenvironment{fig}{\begin{figure}[tp]\begin{minipage}{\textwidth}\centering}{\end{minipage}\end{figure}}

\graphicspath{{./fig/}}

\begin{document}
\title{Cosmic rays of leptons from Pulsars and Supernova Remnants}
\author{Roberto A. Lineros\footnote{email:\texttt{lineros@to.infn.it}} }
\address{INFN, sezione di Torino, I-10122 Torino, Italy and Dipartimento di fisica teorica, Universit\`a di Torino, I-10122 Torino, Italy.}

\begin{abstract}
The latest results from PAMELA and FERMI experiments confirm the necessity to improve theoretical models of production and propagation of galactic electrons and positrons. 
There are many possible explanations for the positron excess observed at energies larger than 10~GeV and for some features around 1~TeV in the total flux of electrons and positrons. 
Supernovae are astrophysical objects with the potential to explain these observations. 
In this work, we present an updated study of the astrophysical sources of lepton cosmic rays and the possible and the possible explanation of the anomalies in terms of astrophysical sources.
\end{abstract}

\keywords{cosmic rays, astrophysical sources, supernovae, pulsars}

\bodymatter
\begin{flushleft}
  {\scriptsize Proceeding for the Conference on Cosmic Rays for Particle and Astroparticle Physics,  CRICATPP 2010, Como, Italy. \\}
  {\scriptsize Preprint DFTT 20/2010 }
\end{flushleft}

\section{Introduction}
During the last decades,  cosmic ray experiments like HEAT~\cite{1997ApJ...482L.191B}  and AMS01~\cite{2000PhLB..484...10A} have found very interesting features in the cosmic rays measuremnts, specially in the component of \emph{electrons}\footnote{in this work, we refer to electrons and positrons as \emph{electrons}} . 
The theoretical prediction of the positron fraction (number of positron divided by the number of \emph{electrons} per unit of energy) based on standard sources mismatched the experimental results.
This feature corresponds to an increment in the fraction for energies larger than 10~GeV.
Recently, newer and refined experiments like PAMELA~\cite{Adriani:2008zr} and FERMI~\cite{2010arXiv1008.3999F, 2010arXiv1008.5119T} have presented more accurate measurements of the positron fraction and the \emph{electrons} flux, respectively, confirming the positron excess and other anomalies present in the observations.
Let us point that these anomalies have not been completely explained by standard theoretical predictions.\\

In this work, we analyze the potential of astrophysical sources (supernova remnants, pulsars, and secondary \emph{electrons}) to explain these anomalies. 
Moreover, we consider the effect of uncertainties in the propagation model and in the sources to verify the robustness of a possible discovery.\\
\section{Propagation of electrons and positrons}
Since \emph{electrons} are injected in the interstellar medium, these cosmic rays are affected by the inhomogeneous galactic magnetic field, radiation fields from stellar activity and from the microwave background, and the interstellar gas.  
The continuous interaction with the environment makes them to loose (or gain) energy and to propagate following a diffusion pattern.\\

The model of propagation of \emph{electrons} is based on a continuity equation for the number density of cosmic rays, known as transport equation, which contains most of the physical processes that those have to face:
\begin{eq}
  \label{eq:std_te}
  \frac{\partial \psi}{\partial t}  - \nabla \big( D(\ener{}) \nabla \psi \big) - \frac{\partial}{\partial \ener{}} \big(b(\ener{})
\psi \big) = q(\vec{x},\ener{}) \, ,
\end{eq}
where $\psi$ is the number density of electrons or positrons per unit of energy, $D(\ener{})$ is the diffusion term, $b(\ener{})$ the energy loss term, and $q(\vec{x},\ener{})$ is the source term.\\

The diffusion term is usually considered homogeneous and isotropic in space with a power law dependence in energy:
\begin{eq}
  D(\ener{}) = K_0 \, \left(\frac{\ener{}}{\ener{0}}\right)^{\delta} \, ,
\end{eq}
where $K_0$ and $\delta$ are phenomenological parameters inspired by magnetohydrodynamics models of the interstellar medium. 
The normalization energy scale $\ener{0}$ here is fixed at the value of 1~GeV.\\

The energy loss term depends on the interaction between the galactic environment and the type of cosmic rays.
The process that dominates the energy evolution of \emph{electrons} at the GeV--TeV scale is the inverse Compton scattering with the Interstellar Radiation Field and the synchrotron emission related to the Galactic magnetic field.\\

For simplicity, the inverse Compton scattering can be described by the Thomson regime which is valid for \emph{electrons} with energy lower than a certain threshold  (details in Refs.~\refcite{2010arXiv1002.1910D} and~\refcite{1970RvMP...42..237B}). 
In this regime, the energy loss term corresponds to:
\begin{eq}
b(\ener{}) = \frac{\ener{0}}{\tau_E} \left(\frac{\ener{}}{\ener{0}}\right)^2 \, , 
\end{eq}
where $\tau_E$ is the energy loss scale time which is calculated from the total photon energy density.\\ 
At higher energies, the Thomson regime is not longer valid to describe the interaction between \emph{electrons} and the photon bath.
The Klein-Nishina regime~\cite{springerlink:10.1007/BF01366453} becomes more accurate than the Thomson regime for the interaction between GeV--TeV \emph{electrons} with the ultraviolet, infrared and microwave component of the interstellar radiation {field~\cite{1970RvMP...42..237B, 2010arXiv1002.1910D}}.\\

For generic functions of the diffusion and energy loss terms,  solutions of the transport equation~(Equation \ref{eq:std_te}) are fully analytical. 
We highlight the solutions for continuous and burst injection cases (details in Refs.~\refcite{2008arXiv0812.4272L} and~\refcite{2010arXiv1002.1910D}) because those are fully of physical meaning.
We remark that the analytical approach allows us to study the importance of different sources and to scan the propagation space of parameters in rather short time compared to numerical methods.\\

\subsection{Uncertainties}
Most of the processes included in the propagation models present theoretical and observational uncertainties.
The principal propagation parameters are the diffusion parameters \mbox{($K_0$, $\delta$)}, and the half-thickness ($L_z$) of the propagation {zone~\cite{1980Ap&SS..68..295G}}.
This zone corresponds to a cylinder of radius equal to the galactic one~(20 kpc) and half-thickness $L_z$. 
The model also includes the escape of cosmic rays from the propagation zone by imposing that cosmic rays density vanishes at the boundaries. 
One method to constrain the propagation space of parameter is based on observation of the Boron/Carbon ratio {(B/C)~\cite{2001ApJ...555..585M}}. 
In principle, this method is less sensitive to injection properties of the source, allowing a better determination of propagation parameters.
For the scope of our analysis, we assume that propagation space of parameters is common for all species of cosmic rays.\\

\section{Astrophyscial sources of \emph{electrons}}
In the GeV--TeV range of energy, \emph{electrons} can be produced in many different ways. 
We highlight the astrophysical sources: spallation with the interstellar medium (secondaries), and supernovae.\\

\subsection{Secondaries}
This component is the result of the spallation of nuclear cosmic rays (protons and alpha particles) off the interstellar gas, which is mainly composed by hydrogen and helium.\\
The environmental factors that determine the flux of secondaries are the amount of interstellar gas, the cosmic ray density, and the production rate of \emph{electrons} related to the nuclear scattering.
Moreover, secondary electrons and positrons are produced in similar amounts. 
The secondary contribution, in the case of electrons, is not dominant compared to the observed flux.
%
On the other hand, secondary positrons encompass the low energy positron flux~\cite{2009A&A...501..821D}. 
We highlight that the compatibility among observation and propagation space of parameter reaffirm the idea that propagation models are common for every specie of {cosmic rays~\cite{2009A&A...501..821D}}. 
However, the propagation uncertainties produce bigger uncertainties in the positron flux than in the case of nuclear cosmic rays calculations.\\

\subsection{Supernova remnants and pulsars}
Supernovae are the astrophysical objects responsible of most of the galactic cosmic rays.
Moreover, those may give birth to two astrophysical objects with potential to produce \emph{electrons}: Supernova remnants and pulsars.\\
Supernova remnants can expel a big fraction of the electron cosmic rays which is contained inside the former star, but very few positrons. 
Different mechanisms have been proposed to enhance the positron {production~\cite{2009PhRvL.103e1104B}}, however, it seems to be not enough to explain the anomalies.
On the other hand, pulsars are able to produce in same amount positrons and electrons due to the interaction of pulsar's magnetic field with ambient photons. \\

In both cases, the injection spectra is expected to follow a power-law like function:
\begin{eq}
  q_{\textnormal{SN}} \propto Q_{0} \ener{}^{-\gamma} \exp\left(- \frac{\ener{}}{\ener{c}} \right) ,
\end{eq}
where $Q_{0}$ is the injection intensity of electrons/positrons, which is calculated from the averaged energy released by the supernova (or pulsar) and the efficiency to convert this energy into electron kinetic energy (details in Ref.~\refcite{2010arXiv1002.1910D}), $\gamma$ is the power index, and $\ener{c}$ is a cut-off energy ($\sim$1--10~TeV) suggested by FERMI and {HESS~\cite{2008PhRvL.101z1104A}}.\\

Let us stress that supernovae are distributed \mbox{non-smoothly} in space and time.
The diffusive behavior of propagated \emph{electrons} smooths any evidence of discrete source distribution at low energy, because this range is dominated by the older and farther supernovae.
On the other hand, it is expected to be observed some features at high energies due to the contribution of younger and closer objects.\\

In figure~\ref{fig3}, we present some of the results of a detailed study of astrophysical sources of \emph{electrons} at the GeV--TeV {scale~\cite{2010arXiv1002.1910D}}.
We remark that under conservative assumption of injection profiles, the observations by FERMI and PAMELA can be encompassed with just the astrophysical component i.e. supernova remnants, pulsars and secondaries.
Also, the presence of inhomogeneities in the local source distribution naturally explains the features observed by FERMI at the TeV range. \\

\begin{fig}
\centering
\resizebox{\hsize}{!}{\includegraphics{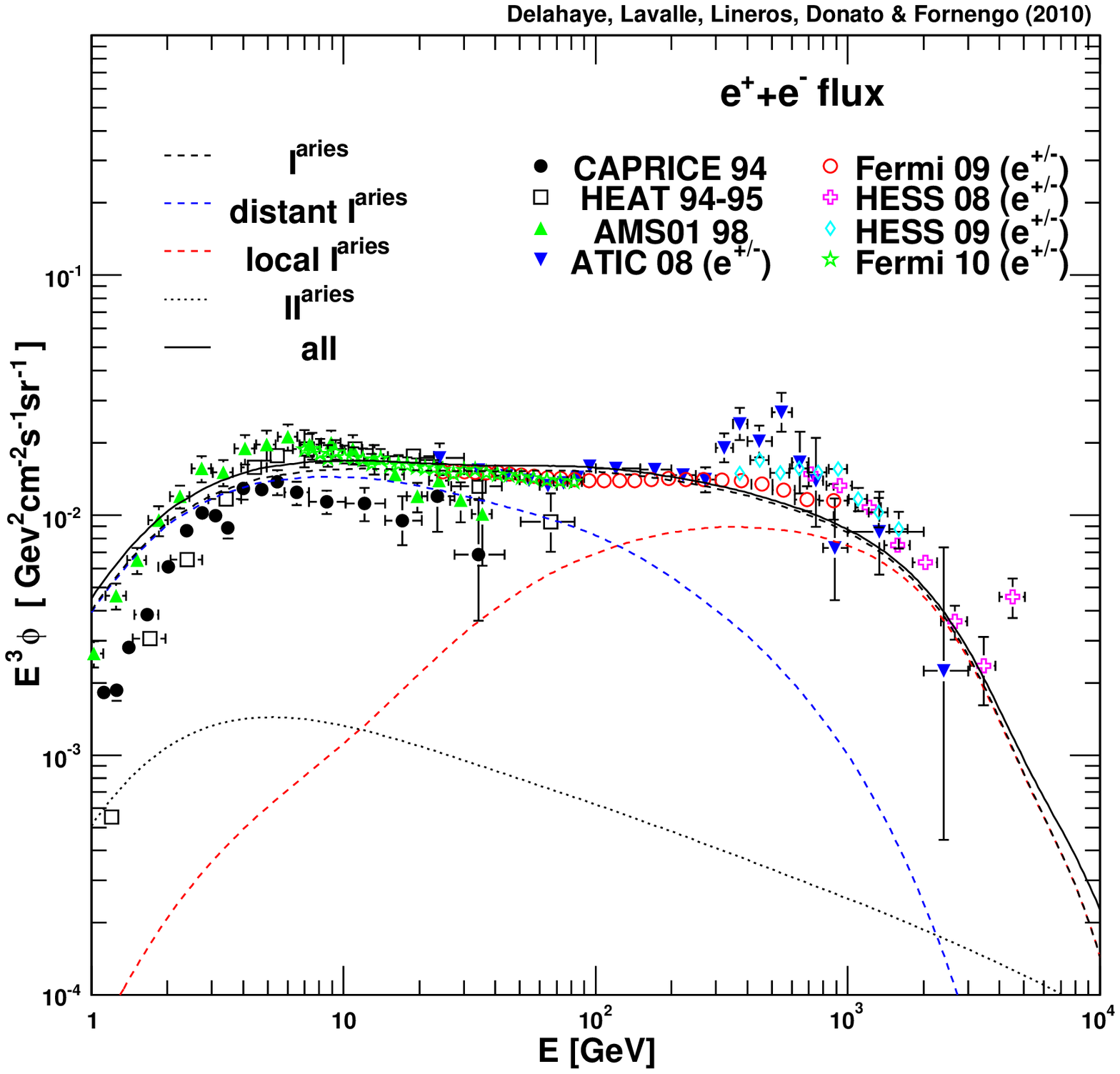}\includegraphics{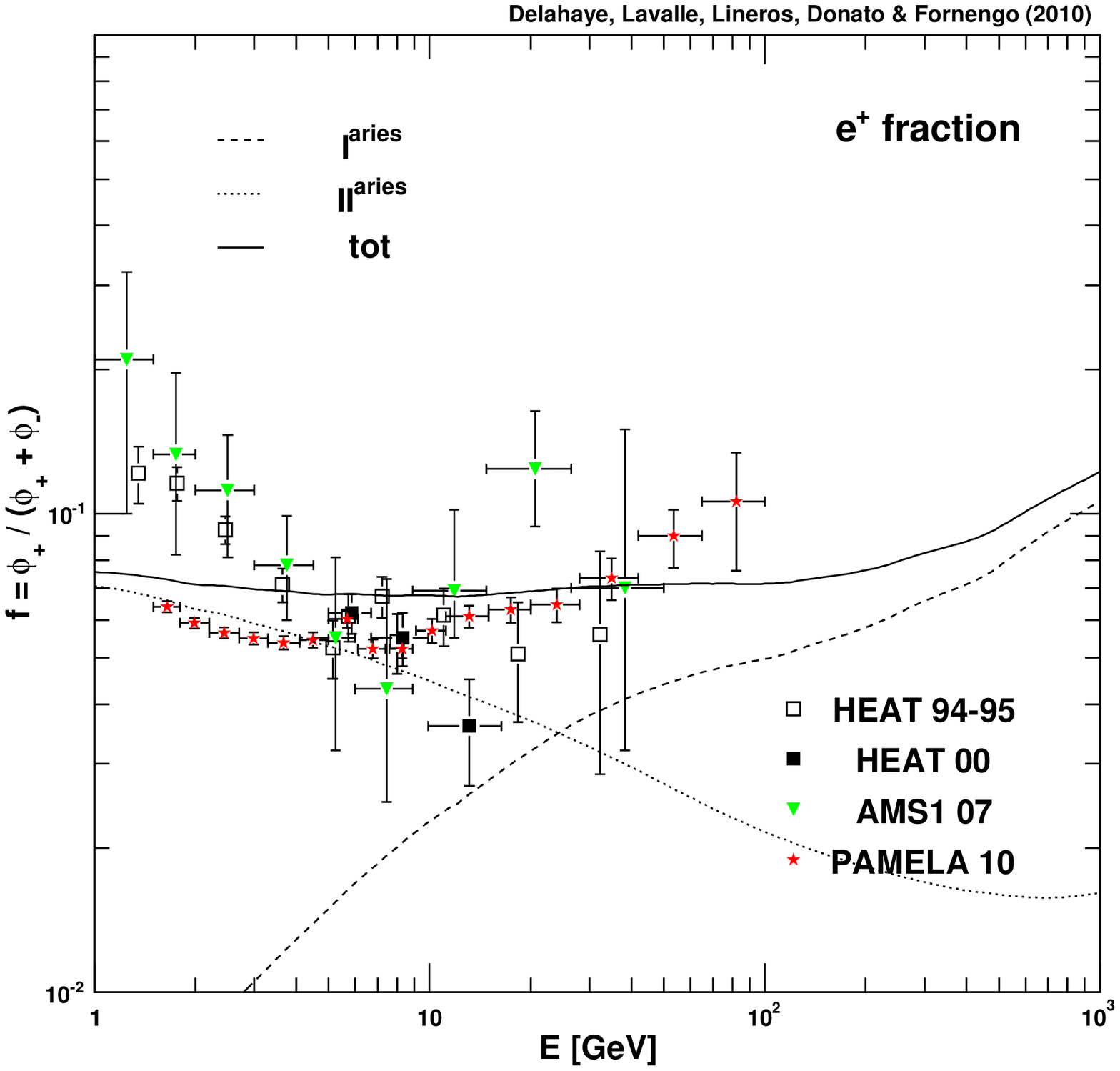}}
\caption{\label{fig3} 
\emph{Electrons} flux (left) and positron fraction (right) versus energy. 
\emph{Electrons} are calculated using information about supernova remnants, pulsars, and also contribution from secondaries.
Both observables are encompassed by the conservative assumption about astrophysical sources.
Further details in Ref.~\refcite{2010arXiv1002.1910D}.}
\end{fig}

\section{Conclusions}
The cosmic rays anomalies observed by PAMELA and FERMI, have triggered a revolution in our understanding about the galactic environment.
Dark matter is far the most exiting solution to the puzzle, but this supposition considers that the astrophysical background is absolutely known.
Uncertainties in the propagation model and in the sources make harder the task to discriminate a possible non-astrophysical source. 
After some detailed study regarding galactic supernova population, it seems natural that supernova remnants and pulsars may be the solution to these anomalies.
It is indispensable to refine the propagation model by using other complementary observables like diffuse gamma emission and radio observations.

\section*{Acknowledgements}
R.L. is grateful to T. Delahaye, J. Lavalle, N. Fornengo, P. Salati, and F. Donato for the rich collaboration and discussion at different stages of the publications referred in this proceeding. Also, R.L. acknowledges financial support given by Ministero dell'Istruzione, dell'Universit\`a e della Ricerca (MIUR), by the University of Torino (UniTO), by the Istituto Nazionale di Fisica Nucleare (INFN) within the Astroparticle Physics Project, and by the Italian Space Agency (ASI) under contract Nro: I/088/06/0. 

\bibliographystyle{ws-procs9x6}
\bibliography{bib/lineros-CRICATPP}
\end{document}